\documentclass{PoS}
\usepackage{graphicx}
\newcommand{\be}{\begin{equation}}
\newcommand{\ee}{\end{equation}}

\title{Universal properties of large $N$ phase
transitions in Wilson loops}

\ShortTitle{Phase transitions in Wilson loops}

\author{R. Narayanan
\\Department of Physics, Florida International University, Miami,
FL 33199, USA\\E-mail: \email{rajamani.narayanan@fiu.edu}}
\author{\speaker{H. Neuberger}
\\ Rutgers University, Department of Physics and Astronomy,
Piscataway, NJ 08855, USA\\E-mail: \email
{neuberg@physics.rutgers.edu}}

\abstract{Numerical studies support the conjecture that 
in continuum planar QCD the eigenvalue density of a Wilson loop 
operator undergoes a transition as the loop is dilated while keeping 
the loop shape fixed. A second part of the conjecture is that the 
transition obeys large $N$ universality and that this universality 
class is the same in 2, 3 and 4 Euclidean space-time dimensions. 
The focus of the talk will be on clarifying precisely 
what the conjecture is claiming.}

\FullConference{The XXV International Symposium on Lattice Field Theory\\
         July 30-4 August 2007\\
         Regensburg, Germany}

\begin{document}

\section{Introduction.}

The main claim of this talk is the following:
The eigenvalue distribution associated with a circular, 
planar Wilson loop at infinite $N$ undergoes a phase transition 
of universal character as the loop is dilated.
The phase transition occurs when the eigenvalue distribution 
changes from having support on an arc with two 
distinct endpoints to covering the entire unit circle.
The transition persists  in the continuum limit for properly 
regulated Wilson loops, $W$. For these operators, the short 
distance -- long distance cross-over occurs over a scale 
range that shrinks to zero as $N$ increases~\cite{nn}.

\section{Universality.}

At infinite $N$, for a small loop, 
the tail of the distribution at eigenvalues close to $-1$  disappears, 
and there is a gap. For finite $N$ the tail is suppressed as $e^{-cN}, c>0$.
For a large loop there is no gap. The transition happens at a critical scale
where it is governed by a large $N$ universality class. The universality class 
is the same in two, three and four Euclidean dimensions. 
The two dimensional model is exactly soluble and therefore the universality class
is fully understood. 

\section{The exactly soluble representative of 
the large $N$ universality class.}

A general formulation of the universality class~\cite{janik} is in 
terms of a random multiplicative ensemble: 
Consider $n$ independent $SU(N)$ matrices {\sl i.i.d.} with a measure $d\mu$.  
The Wilson loop $W$ of unit-less area $t=n\epsilon^2$ is represented by a 
product of the unitary matrices, 
in any fixed order, and evolves in $t$ by diffusion on $SU(N)$ starting from a 
point source. This is  exactly true in two dimensions and the
relevance of the associated  Durhuus-Olesen~\cite{do}
universality class to higher dimensions is an exact version 
of the old conjecture of ``dimensional reduction''.

\section{Average characteristic polynomial.}

The average characteristic polynomial $Q_N$ of the fluctuating unitary
matrix $W$ at a unit-less loop scale $t$ can be defined in any dimension and captures 
the essential features of the collective behavior of the 
$W$ eigenvalues:
\be
Q_N(z,t)\equiv \langle \det(z-W(t)) \rangle
\ee

We are interested in the region around 
eigenvalue -1 which affects $Q_N(z,t)$ mainly around $z=-1$. In two dimensions 
the exact result is obtained as $\epsilon\to 0,~n\to\infty$ 
at fixed $t$ and is critical at $t=t_c=4$.

\subsection{$Q_N(z,t)$ in two dimensions.}

\begin{figure}
\vspace{1.1cm}
\begin{center}
\centerline{\includegraphics[width=0.6\textwidth]{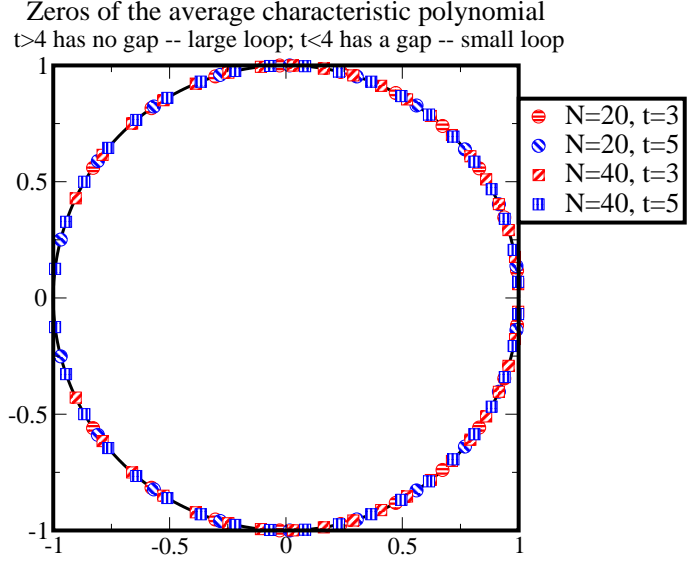}}
\caption{Locations of ``charges'' in the complex $z$-plane; 
we are interested in a small arc centered at $z=-1$.\label{zeros}}
\end{center}
\end{figure}

In two dimensions we can calculate $Q_N (z,t)$ exactly. Applying the 
Lee-Yang theorem to the result
we prove that all the zeros of $Q_N(z,t)$ lie on the unit circle.
Figure~\ref{zeros} shows this graphically. 

\subsection{``Double scaling limit'' at large $N$ in two dimensions.}

At infinite $N$ the logarithmic derivative of $Q_N$ is a 
two dimensional electric field associated with a static charge 
distribution representing the 
eigenvalues of $W$; for one realization of $W$
each eigenvalue is counted as a unit charge located 
at the value of the eigenvalue. When there 
is no gap in the eigenvalue distribution
there is a jump in the field at -1 on the unit circle.
When there is a gap there is no jump. The jump gets smoothed out at finite
$N$ in a universal manner. 

\begin{figure}
\vspace{1.1cm}
\begin{center}
\centerline{\includegraphics[width=0.6\textwidth]{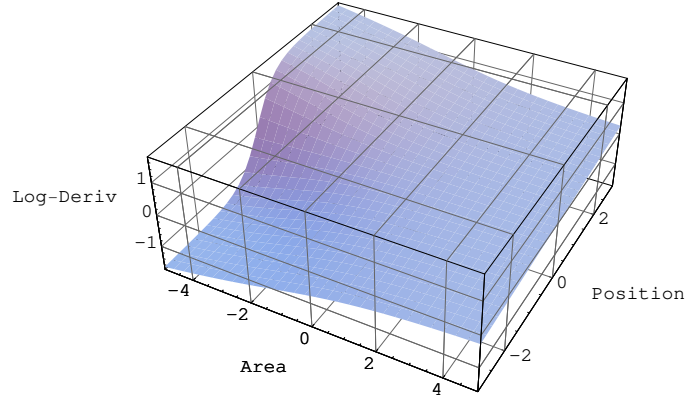}}
\caption{Plot of the logarithmic derivative of $\zeta$ with
respect to $\xi$ for different values of the ``area'' ($\alpha$)
and ``position'' ($\xi$). One sees the smooth remnant
of the singularity at $\xi=0$ and its dependence on the area.\label{smoothfig}}
\end{center}
\end{figure}

In terms of the characteristic polynomial one captures 
the smoothing of singularities by going to a special
limit, in which two new parameters are 
scaled by $N^{\nu_i}$ and kept fixed as $N$ goes to
infinity. The region around $z=-1$ and $t=t_c=4$ gets infinitely amplified. 
We restrict our attention to 
the real $z$ axis. One introduces the
variables $\xi$ and $\alpha$:
\be
z=-e^y,~~~\frac{1}{t}=\frac{1}{4}+a
\ee
with
\be
y=\frac{2}{6^{1/4} N^{3/4}}\xi,~~~a=\frac{1}{4\sqrt{3N}} \alpha
\ee
We now take $N$ to infinity keeping the variables $\xi$ 
and $\alpha$ at fixed finite values. The numerical prefactors
are chosen to make the limit of $Q_N(z,t)$ simple.
Up to non-universal factors, $Q_N(z,t)$ 
approaches a function $\zeta(\alpha,\xi)$:
\be
\zeta(\alpha,\xi)\equiv\int du e^{-u^4-\alpha u^2 +\xi u}
\ee
$\partial\log \zeta(\alpha,\xi)/\partial \xi$  
is similar to a two dimensional electric field and its discontinuity across
a charged line is smoothed as the line charge density 
gets depleted in the limit. Figure~\ref{smoothfig} shows the smooth result.  

The meaning of large $N$ universality is that 
an evaluation of $Q_N(z,W(t))$ also in three or
four dimensional $SU(N)$ pure Yang-Mills theory,
where $t$ labels a dilation parameter and the shape of the loop is held fixed,
would asymptote to $\zeta(\alpha,\xi)$ 
with the parameters $\alpha$ and $\xi$ defined from
$y$ and $t-t_c$ using $N^{1/2}$ and $N^{3/4}$
multiplied by some finite numerical amplitudes dependent on the theory 
and on the shape of the Wilson loop. 
The essence of the claim is that the asymptotic 
behavior will be describable 
by only two parameters and that the 
exponents associated with the factors of $N$
relating $\alpha,\xi$ to $t,y$ will be exactly $\nu_{1,2}=1/2,3/4$. 

\subsection{Convergence to infinite $N$ in two dimensions.}

\begin{figure}
\vspace{1.1cm}
\begin{center}
\centerline{\includegraphics[width=0.6\textwidth]{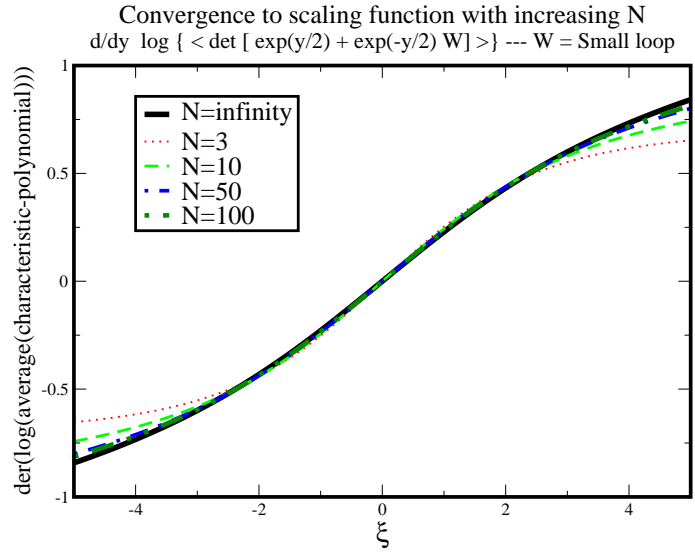}}
\caption{Wilson loop with a gap.\label{Gapped}}
\end{center}
\end{figure}

\begin{figure}
\vspace{1.1cm}
\begin{center}
\centerline{\includegraphics[width=0.6\textwidth]{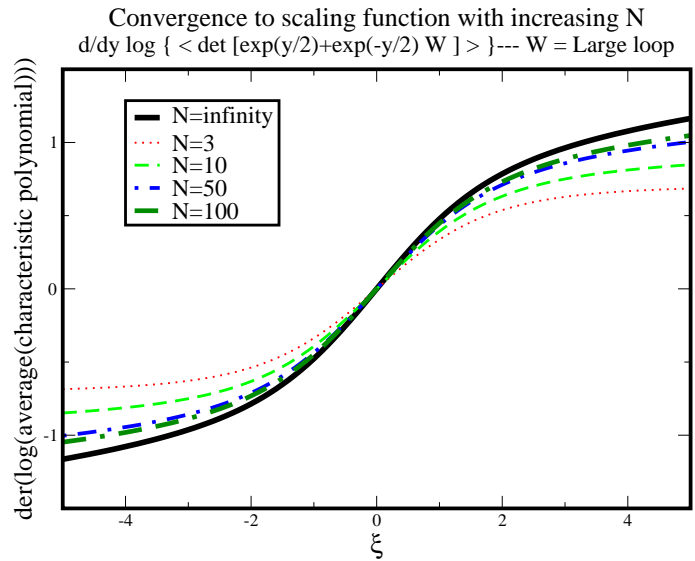}}
\caption{Wilson loop without a 
gap.\label{Ungapped}}
\end{center}
\end{figure}

To test this claim in simulation we first need an indication for how
high $N$ ought to be for the universal behavior to set in.
We plot two cross-sections 
through the surface of Figure~\ref{smoothfig} and check how rapidly they
are approached as $N$ increases in two dimensions, where we have exact formulae
also at finite $N$. The result is shown in
Figures~\ref{Gapped} and~\ref{Ungapped}, for a Wilson loop 
that at infinite $N$  would have a gap and one that would not.
For large loops the limit is approached slowly.

\section{Parameter extraction by nonlinear three parameter fits.}

In view of the slow approach to the asymptotic form, we 
ask how least-square 
best fits of it 
to finite $N$ expressions in two dimensions would approximate the
values of the critical scale $t_c$ and 
the two amplitudes relating $y$ and $t-t_c$
to $\xi$ and $\alpha$, when we assume $\nu_i$ to be fixed at $1/2$ and $3/4$. 
The parameter corresponding to $t_c$ is denoted by $A3$ and the numerical
coefficients multiplying $N^{\nu_i}$ entering $\alpha$
and $\xi$ are denoted by $A2$ and by $A1$ respectively. In two 
dimensions the correct values are $A1=A2=A3=1$.

\subsection{Two dimensions.}
\begin{figure}
\vspace{1.1cm}
\begin{center}
\centerline{\includegraphics[width=0.6\textwidth]{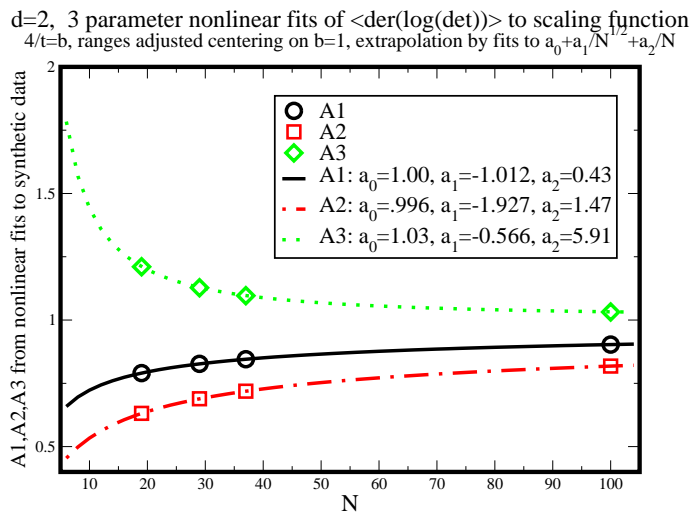}}
\caption{Analysis of synthetic two dimensional data.\label{twodsynt}}
\end{center}
\end{figure}

Figure~\ref{twodsynt} shows the result of doing global least square fits 
to $\partial\log\zeta(\alpha,\xi)/\partial \xi$ of 
synthetic data obtained from the exact solution in two dimensions. 
As indicated in the figure, one extrapolates to infinite $N$ by fitting to
a series in $1/\sqrt{N}$ -- this form is a consequence of $\nu_{1,2}=1/2,3/4$.

This exercise shows that one may 
get close to the right answer even from values of
$N$ that are practical, but for 
accuracy at the level of a fraction of a percent one would need to go
to $N=100$. This value is too high for practical Monte Carlo simulations
in three or four dimensions. However, $N\sim 50$ is practical and sufficient for few percent accuracies.

\subsection{Three dimensions.}

\begin{figure}
\vspace{1.1cm}
\begin{center}
\centerline{\includegraphics[width=0.6\textwidth]{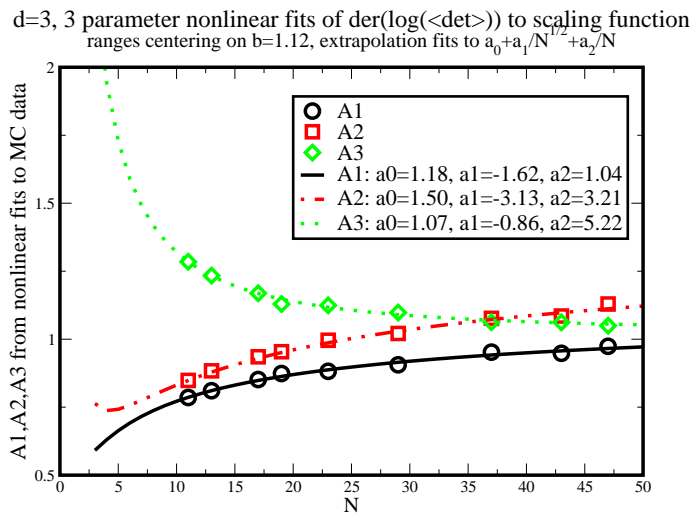}}
\caption{Analysis of three dimensional Monte Carlo data.\label{threedmc}}
\end{center}
\end{figure}

We now turn to three dimensions 
and carry out test runs at some convenient point.
Figure~\ref{threedmc} shows the results.
We choose a case where the critical Wilson loop size 
occurs at an inverse 't Hooft 
coupling $b$ close to unity, so we can just use definitions
identical to those in our analysis of the synthetic two dimensional data
by identifying the parameter $t$ there with $\frac{4}{b}$ here. This is 
true only to 10 percent, but is enough to give us a feel for the order of
magnitude of the non-universal amplitudes $A1,A2$ in three dimensions.

Unlike in two dimensions, the data has 
statistical errors and this further impacts accuracy. 
Nevertheless, one obtains an indication that universality holds with 
amplitudes numerically quite close to their values in two dimensions.
This is accidental to some degree 
and cannot hold in general, as in three
dimensions there are Wilson loop of shapes unattainable in two dimensions. 

\section{Conclusions.}
We conclude that it should be possible to check 
our main claim in three dimensions but 
the method of analysis ought to be improved. 
So far, the numerical results in 3D are in agreement with large $N$ universality.
Amplitudes relating to simple square Wilson loops
in three dimensions come out surprisingly close 
to their two dimensional values; perhaps, this is a reflection of the 
approximate dimensional reduction known to hold since the 80's.

\acknowledgments

R.N. acknowledges partial support by the NSF under grant number
PHY-055375. 
H. N. acknowledges partial support
by the DOE under grant number
DE-FG02-01ER41165.

\end{document}